\documentstyle[12pt]{article}
\def\Re{{\cal R \mskip-4mu \lower.1ex \hbox{\it e}\,}}
\def\Im{{\cal I \mskip-5mu \lower.1ex \hbox{\it m}\,}}
\def\snu{\ifmmode \tilde\nu \else $\tilde\nu$\fi}
\def\slep{\ifmmode \tilde l \else $\tilde l$\fi}

\def\sub#1{_{\lower.25ex\hbox{$\scriptstyle#1$}}}
\def\sul#1{_{\kern-.1em#1}}
\def\sll#1{_{\kern-.2em#1}}
\def\sbl#1{_{\kern-.1em\lower.25ex\hbox{$\scriptstyle#1$}}}
\def\ssb#1{_{\lower.25ex\hbox{$\scriptscriptstyle#1$}}}
\def\sbb#1{_{\lower.4ex\hbox{$\scriptstyle#1$}}}

\def\to{\rightarrow}
\def\mh{\ifmmode m\sbl H \else $m\sbl H$\fi}
\def\mch{\ifmmode m_{H^\pm} \else $m_{H^\pm}$\fi}
\def\mt{\ifmmode m_t\else $m_t$\fi}
\def\mc{\ifmmode m_c\else $m_c$\fi}
\def\mz{\ifmmode M_Z\else $M_Z$\fi}
\def\mw{\ifmmode M_W\else $M_W$\fi}
\def\mws{\ifmmode M_W^2 \else $M_W^2$\fi}
\def\mhs{\ifmmode m_H^2 \else $m_H^2$\fi}
\def\mzs{\ifmmode M_Z^2 \else $M_Z^2$\fi}
\def\mts{\ifmmode m_t^2 \else $m_t^2$\fi}
\def\mcs{\ifmmode m_c^2 \else $m_c^2$\fi}
\def\mchs{\ifmmode m_{H^\pm}^2 \else $m_{H^\pm}^2$\fi}
\def\ztwo{\ifmmode Z_2\else $Z_2$\fi}
\def\zone{\ifmmode Z_1\else $Z_1$\fi}
\def\mtwo{\ifmmode M_2\else $M_2$\fi}
\def\mone{\ifmmode M_1\else $M_1$\fi}
\def\tb{\ifmmode \tan\beta \else $\tan\beta$\fi}
\def\xw{\ifmmode x\sub w\else $x\sub w$\fi}
\def\ch{\ifmmode H^\pm \else $H^\pm$\fi}
\def\lum{\ifmmode {\cal L}\else ${\cal L}$\fi}
\def\inpb{\ifmmode {\rm pb}^{-1}\else ${\rm pb}^{-1}$\fi}
\def\infb{\ifmmode {\rm fb}^{-1}\else ${\rm fb}^{-1}$\fi}
\def\epem{\ifmmode e^+e^-\else $e^+e^-$\fi}
\def\ppb{\ifmmode \bar pp\else $\bar pp$\fi}

\newskip\zatskip \zatskip=0pt plus0pt minus0pt
\def\matth{\mathsurround=0pt}

\def\atversim#1#2{\lower0.7ex\vbox{\baselineskip\zatskip\lineskip\zatskip
  \lineskiplimit 0pt\ialign{$\matth#1\hfil##\hfil$\crcr#2\crcr\sim\crcr}}}

\renewcommand{\thefootnote}{\fnsymbol{footnote}}

\renewcommand{\arraystretch}{1.5}

\begin{document} \begin{titlepage}
\setcounter{page}{1}
\thispagestyle{empty}
\rightline{\vbox{\halign{&#\hfil\cr
&UMD-PP-93-90\cr
&ANL-HEP-PR-92-123\cr
&OITS-503\cr
&December 1992\cr}}}
\vspace{1in}
\begin{center}

{\Large\bf Radiative $W$ and $Z$ Decays and Spontaneous R-Parity Violation}

\medskip

\normalsize R.N. MOHAPATRA
\\ \smallskip
Department of Physics, University of Maryland, College Park, MD 20742\\
\smallskip
and\\
\smallskip
\normalsize T.G. RIZZO
\\ \smallskip
High Energy Physics Division, Argonne National Laboratory, Argonne, IL 60439\\
\smallskip
and\\
\smallskip
Institute of Theoretical Science, University of Oregon, Eugene, OR 97403\\

\end{center}

\begin{abstract}

We point out that in a class of supersymmetric models where R-parity
violation is induced by the spontaneous breaking of local $B-L$ symmetry,
the R-parity violating $W$ decay $W\rightarrow\slep\gamma$ and
$Z$ decay $Z\rightarrow\snu\gamma$, forbidden in the
minimal supersymmetric standard model (MSSM), occur at an enhanced rate
compared to other models with R-parity breaking. We find that the branching
fractions for these modes can be of order $10^{-5}$ .

\end{abstract}

\renewcommand{\thefootnote}{\arabic{footnote}} \end{titlepage}

{\bf \underline {I.Introduction}}

  Supersymmetrization of the standard model brings along with it the
unpleasant feature that baryon and lepton numbers ($B$ and $L$) are no
longer automatic symmetries of
the Lagrangian. It is therefore customary to impose these symmetries
on the model in order to avoid rapid proton decay or lepton number
violation which are not yet observed in nature. Both these
symmetries are however simultaneously obeyed if a discrete R-parity
symmetry defined as $(-1)^{2S+3B+L}$ is imposed on the Lagrangian.
The additional assumption of R-parity conservation severely limits
the possible interactions among the fermions and their superpartners
in the minimal supersymmetric standard model(MSSM). It not only implies that
all superpartners of standard model particles must be produced in
pairs but also that the lightest superparticle (the neutralino)
must be stable. As mentioned above there is no a priori theoretical
reason for either of these commonly made assumptions to hold.

  Supersymmetric theories without R-parity conservation
were introduced nearly ten years ago[2,3] in order to examine the
experimental constraints on the extent of departure from exact
R-conservation. Two classes of theories were considered: one, where
the R-parity violation is spontaneous[2] and a second, where it is
explicit in the original superpotential[3]. This latter possibility arises
since the symmetries of the conventional gauge interactions of the MSSM do
not {\it a priori} forbid such terms in the superpotential. Many implications
and tests of these two ideas have been subsequently analyzed in literature[4].

  It has recently been noted [5,6] that if the sleptons are lighter
than the $W$ and $Z$ bosons then R-parity violation implies a new
decay channel for the latter, $W\rightarrow\slep\gamma$ and
$Z\rightarrow\snu\gamma$, with the sleptons possibly decaying subsequently to
two quarks via the mediation of R-violating interactions. These new
decay modes of $W$ and $Z$ would be detectable in collider experiments
provided the corresponding branching fractions
are significantly large. It turns out, however[5,6], that in most models with
explicit R-parity breaking
these branching fractions are found to be of order $10^{-7}$ or $10^{-8}$
keeping them beyond the reach of present experiments.

In this note, we focus our attention on a different class of models for
R-violation. It was noted some time ago [7] that in extensions of the
supersymmetric standard
model where the gauge symmetry contains local $B-L$ as an explicit
subgroup, R-parity invariance is automatic. Its violation therefore
can emerge if the spontaneous breaking of $B-L$ is caused by a
non-zero vev for the right-handed sneutrino (i.e., $<\snu^c>\neq 0$).
An additional advantage of this model is that while lepton number
violating terms are induced after spontaneous breaking, baryon number
remains an exact symmetry, thereby avoiding any danger of rapid
proton decay.
It is the purpose of this note to point out that the single photon radiative
$W$ and $Z$ decays are enhanced by two to three orders of magnitude in
these models
compared to the explicit R-parity violating scenarios and there is therefore
a chance that they may be observable in collider experiments.

\smallskip

{\bf \underline{II.Details of the Model}}

    The simplest gauge group which contains the standard model as well as a
$U(1)_{B-L}$ factor is $SU(2)_L\times U(1)_{I_{3R}}\times U(1)_{B-L}$
which is itself in turn a subgroup of the left-right symmetric gauge
group $SU(2)_L\times SU(2)_R\times U(1)_{B-L}$. We will illustrate
our point using the first gauge group which is simpler to analyze although
our results hold as well for the left-right symmetric SUSY model. The
matter spectrum of the model consists of three generations
of quarks and leptons as in the standard model plus three right-handed
neutrinos (denoted by $\nu^c$). Their assignments under the extended
gauge group are given in Table I.
\newpage
\begin{center}
{\bf Table I}
\end{center}
\begin{tabular}{|c||c||c|}\hline
Matter Fields  &  $SU(2)_L\times U(1)_{I_{3R}}\times U(1)_{B-L}$ Quantum
numbers
\\    \hline
  $ Q\equiv (u,d) $ &  ($2, 0, {1\over 3}$ ) \\
  $ u^c$            &  ($1,-{1\over 2},-{1\over 3}$)  \\
  $d^c$             &   ($1,+{1\over 2},+{1\over 3}$)  \\
  $L\equiv (\nu,e)$               &  ($2,0,-1$) \\
  $e^c$             &  ($1,+{1\over2},+1$)  \\
  $\nu^c$           &  ($1,-{1\over 2},1$)   \\
  $H_u$             &  ($2,+{1\over 2},0$)   \\
  $H_d$             &  ($2,-{1\over 2},0$)  \\
\hline
\end{tabular}

  The superpotential for this model can be written as follows:
\begin{eqnarray}
W = & h_l LH_{d}e^c + h_{\nu}LH_u\nu^c + h_u QH_u{u^c} + h_d QH_d{d^c}
+\mu H_uH_d
\end{eqnarray}
In this equation, we have suppressed the generation indices. Before discussing
the R-parity violation, let us comment briefly on the status
of neutrino masses in this model. It is easy to see that the $h_\nu$
terms in the superpotential will induce Dirac masses for the neutrinos
which could {\it a priori} be of the order of the charged lepton masses. The
simplest way to cure this problem is to introduce two $B-L$ carrying
weak isosinglet fields, $\Delta (1,+1,-2)$ and $\bar{\Delta}(1,-1,+2)$,
and let them acquire nonzero vev's of the order a TeV or more. Then a
new coupling in the superpotential of the form $\nu^c\nu^c\Delta$
will give rise to the familiar see-saw mechanism for the neutrino making their
masses small. Without effecting our final conclusions, we adopt
a simpler approach without the extra $\Delta$ fields. In order to understand
the small neutrino masses in this simplified model, we will set all the
elements of the coupling matrix $h_\nu$ to zero except
$h_{{\nu}_{33}}$, so that both $\nu_e$ and $\nu_\mu$
have zero Dirac masses. The Dirac mass induced for the tau neutrino is
then of order of the tau lepton mass itself; the smallness of the $\nu_\tau$
mass can then be understood by
a 3x3 see-saw mechanism as introduced in Ref.[8] after $\nu^c_{33}$
acquires a nonzero vev. To see this, let us assume that the $B-L$ symmetry
breaking is implemented by $<\nu^c>=v_R$ and the rest of the symmetry
breaking is caused by $<H_u>=\kappa_u$ and $<H_d>=\kappa_d$. It is then
easily seen that the $\nu_{\tau}^c$ mixes with the linear combination
of the gauginos corresponding to the generators $I_{3R}$ and $B-L$. Let us
call this gaugino $\lambda_{\nu^c}$. Then the 3x3 mass matrix corresponding
to the $\nu_{\tau}, \nu_{\tau}^c, \lambda_{\nu^c}$ basis appears as follows:

\renewcommand\arraystretch{1.1}

$$   \left( \begin{array}{ccc}
              0   &  h_{\nu}\kappa_u &  0\\

              h_\nu \kappa_u &  0    &  \tilde {g}v_R \\

              0      &     \tilde {g}v_R  &  m_\lambda \end{array}
\right) $$

 The 33-entry in the above matrix represents the SUSY breaking gaugino
mass term. As was already remarked in Ref.8, this leads to a tiny mass
for the left handed neutrino given by $m_\nu\simeq\left(m_D^2.m_\lambda
\over{{\tilde {g}}^2{v_R}^2}\right)$, where $m_D=h_\nu\kappa_u$ and
$\tilde {g}^2=
(g_R^2+g_{BL}^2)/4$. This double see-saw result shows that in the
expression for the neutrino mass there is an additional suppression
coming from the Majorana mass of the gaugino. In supersymmetric theories,
if this Majorana mass of the gaugino is set to zero at the tree level
it can only arise at the two loop level and therefore can be less
than a GeV. We then find that for $m_D\simeq{1GeV}$,
$m_\lambda\simeq 1GeV$ and $v_R\simeq{10TeV}$ we obtain a value for
$m_{\nu_\tau}\leq 10eV$ which is acceptable from both laboratory
and cosmological considerations. Note further that, as $m_\lambda\rightarrow
{0}$, the tau neutrino mass vanishes.

   It is also worth pointing out that if the other entries in the
coupling matrix $h_\nu$ (entries other than the 33 entry) are
nonzero, one could either invoke the usual see-saw mechanism by
introducing the $\Delta$ and $\bar{\Delta}$ as mentioned before,
or introduce three gauged $U(1)_{B-L}$ symmetries, one per
generation, and give vevs to all three $\snu^c$'s. In the
latter case we will have a generalized see-saw mechanism operating
separately within each generation. We will not elaborate on this possibility
here. In either case, the main result of our paper remains unchanged.

  When $\snu^c$ acquires a nonzero vev, it leads to R-parity violating
interactions in the Lagrangian below the scale $v_R$. It is easily seen that
such terms only induce lepton violating terms in the low energy Lagrangian.
By making the off diagonal terms of $h_{l}$ arbitrarily small, we can
make the theory consistent with the observed bounds on lepton number
conservation. The main goal of our paper, i.e., to demonstrate that in this
class of models the radiative $W$ and $Z$ decays are enhanced compared to
other R-parity breaking models, follows directly from this approach.

   Breaking local $B-L$ symmetry by the mechanism employed in this paper
however implies other constraints on the model which have relevance to the
strength of the radiative decays considered. To see this, note that a nonzero
$h_\nu$ coupling combined with nonzero $v_R$ and $\kappa_u$ imply a
nonzero vev for the tau sneutrino, i.e., $<\tilde {\nu_\tau}>=v_L$. In
models with {\it global}
$B-L$ symmetry breaking by $\tilde {\nu^c}$, there exists a massless
Majoron and astrophysical constraints on the Majoron's properties then
imply that [9] $v_L< 10 - 100 MeV$. On the other hand, in our model there
is no massless Majoron. Thus the only constraint on $v_L$ comes from the
fact that it leads to mixing between [10] the tau neutrino and the $B-L$
gaugino as well as between the tau lepton and the wino.  This mixing can cause
departures from the apparent universality of leptonic decays of the
muon and the tau leptons. These universality constraints are however
weaker than those which apply in the case of the Majoron and are easily
satisfied for $v_L$ less than a few GeV. As
a conservative upper limit for $v_L$, we will choose a value of 1 GeV. In
this case, the tau-gaugino mixing angle is less than 0.01. To see what
constraints are implied by this, we note that
\begin{equation}
v_L \simeq {h_\nu}{v_R}\left({\mu\kappa_d + m_{3\over2}\kappa_u}\over
{M^2}\right)
\end{equation}
For $v_R\simeq 10 TeV$, $M\simeq 100 GeV$, $\kappa_u , \kappa_d\simeq 100 GeV$,
$m_{3\over2}\simeq 100 GeV$ and $h_\nu\simeq 10^{-2}$, the above
bound on $v_L$ is seen to be easily satisfied. Clearly as the constraints
on $v_L$
improve, some of the parameters on the right-hand side of Eq.2 will become
smaller. We will see that the strength of the radiative $W$ and $Z$
decays will depend on the values of these same parameters.

   Let us briefly examine the charged and neutral scalar bosons in this theory.
In the unitary gauge, there are the sleptons of the electron and muon
type, which we assume are unmixed with the usual Higgs bosons. Due
to the fact that $<\tilde {\nu_{\tau}}^c>\neq 0$, the tau slepton mixes with
both $H_u$ and $H_d$ with the result being that we have two charged
physical scalar bosons each of which
will contain an admixture of the $\tilde {\tau}^+$. In the MSSM, where the tau
slepton does not mix with the Higgs bosons, the physical charged Higgs
boson is almost always heavier than the $W$ boson so that the $W$ cannot decay
into it.
In our model, there are two new ways in which the charged
Higgs boson sector is different. First, the usual charged scalar can
mix with the $\tilde {\tau}^+$ as already mentioned above
and, secondly, the mass of $H_u^+$ gets
additional contributions from the nonzero vev of $\snu^c$. Because
of this, the physical eigenstate with the larger mass is predominantly the
MSSM Higgs boson while the other eigenstate is predominantly the tau slepton
which can be lighter than the $W$.

 As we will see below, this lighter eigenstate
will couple predominantly to the $t$ and $b$ quarks. Similarly, there
will also be a neutral Higgs boson which will be predominantly
the tau sneutrino and it will also have significant couplings to
the t and b quarks. We will assume that its mass is also less than the
mass of the $Z$ boson. These two particles can then appear in the radiative
decay of the $W$ and $Z$.

\smallskip

{\bf \underline {III.Calculation of The Single Photon Radiative Decay
of $W$ and $Z$}}

Now we are ready to discuss the induced lepton number violating term that
is responsible for the single photon radiative decays of the massive
gauge bosons. The tree diagram shown in Fig.1 which arises after
supersymmetry breaking leads after $B-L$ breaking to
the effective $tb\slep$ interaction of the form
\begin{eqnarray}
{\cal L} = &  f[~\tilde l_L\bar b(1+\gamma_5)t + \tilde{\nu}_L
\bar t(1+\gamma_5)t~] + h.c.
\end{eqnarray}
where $f$ is given in our model by
\begin{eqnarray}
f = & \left({{h_t}{h_\nu}m_{3\over2}{v_R}}\over{M_{H_u}^2}\right)
\end{eqnarray}
As mentioned before, we have a final state scalar that will be a linear
combination of the charged Higgs fields which is predominantly the tau slepton.

   If we choose $M_{H_u}\simeq 100 GeV$, and all other parameters as
discussed above, we see that $f$ can easily be of order one to three.
We will assume this in making the predictions for the radiative branching
ratios given below. The coupling in Eq.3 induces the decay
amplitude for $W\rightarrow \tilde {\tau}\gamma$ via the one loop
diagram in Fig.2. With this
normalization, the amplitude for the $W\to \slep\gamma$ decay process can be
written as
\begin{equation}
{\cal A}=\Big[F_1{(q_\nu k_\mu-g_{\nu\mu}k\cdot q)\over M_W^2} +
iF_2{\epsilon_{\mu\nu\sigma\tau}q^\sigma k^\tau\over M_W^2}\Big]
\epsilon^\mu_\gamma \epsilon^\nu_W
\end{equation}
with q(k) being the momentum of the photon($W$). In terms of the form factors
$F_{1,2}$, the decay width is given by
\begin{equation}
\Gamma(W\to \slep\gamma)={M_W^3 \over {96\pi^2}}(F_1^2+F_2^2)
\Big(1-{m_{\slep}^2 \over {M_W^2}}\Big)^3
\end{equation}
with $m_{\slep}$ being the slepton mass. Defining the mass difference,
$\delta=m_{\slep}^2-M_W^2$, we find that $F_{1,2}$ can be written as
\begin{eqnarray}
F_1 &={-ie gN_c m_t f \over 4\sqrt {2}\pi^2\delta}
[Q_uI_1+Q_dI_2] \\
F_2 &={-ie gN_c m_t f \over 4\sqrt {2}\pi^2\delta}
[Q_uI_3+Q_dI_4] \nonumber
\end{eqnarray}
where $N_c$=3 is the usual color factor, $m_t$ is the t-quark mass, $g$ is the
conventional weak coupling constant, $Q_{u,d}$ are the electric charges of
the up- and down-quarks, and $I_i$ can be expressed as sums of
parameter integrals:
\begin{eqnarray}
I_1 & = & 1+(2m_t^2 \delta^{-1}-1) G_{-1}(m_t,m_b)
+2[\delta^{-1}(m_b^2-m_t^2-M_W^2)+1/2] G_0(m_t,m_b) \nonumber \\
    & + & 2M_W^2 \delta^{-1} G_1(m_t,m_b) \nonumber \\
I_2 & = & 1+2m_b^2 \delta^{-1} G_{-1}(m_b,m_t)
+2[\delta^{-1}(m_t^2-m_b^2-M_W^2)
-1/2] G_0(m_b,m_t) \nonumber \\
    & + & 2M_W^2 \delta^{-1} G_1(m_b,m_t) \\
I_3 & = & G_{-1}(m_t,m_b)-G_0(m_t,m_b) \nonumber \\
I_4 & = & -G_0(m_b,m_t) \nonumber
\end{eqnarray}
where the $G_n$ are given by
\begin{equation}
G_n(m_i,m_j)=\int^1_0 dz z^n \ln\Bigg[{m_i^2(1-z)+m_j^2z-z(1-z)m_{\slep}^2
\over
m_i^2(1-z)+m_j^2z-z(1-z)M_W^2} \Bigg]
\end{equation}
It is important to note that both $F_{1,2}$ are proportional to $m_t$.

 Similar considerations apply to the $Z$ decay to sneutrino plus photon.
As in the previous case, here also we have an enhancement arising from
the presence of the top-quark mass although a suppression also occurs due the
heavy top-quark propagators in the loop. The decay rate for the radiative
$Z$ decay can be written as:
\begin{equation}
\Gamma(Z\rightarrow \snu\gamma)={{\alpha^2 M_{Z}^3}\over{192\pi^3
m_t^2}}f^2\left(1-{{m_{\tilde {\nu}}^2}\over{M_Z^3}}\right)^3
{{(1-8/3sin^2\theta_W)^2}\over{sin^2\theta_W cos^2\theta_W}} I
\end{equation}
with
\begin{equation}
I=|I_1(x,y)-I_2(x,y)|^2 +| I_2(x,y)|^2
\end{equation}
where the $I_i's$ are complicated functions of $x=(4m_t^2/m_{\snu}^2)$
and $y=(4m_t^2/M_Z^2)$ given in Ref.11.
The predictions for the $W \to \slep \gamma$ and $Z \to \snu \gamma$
branching fractions as functions of the $\snu$ and $\slep$ masses
are given in Figs.3 and 4 assuming $f$=1. We see that they can be as large as
$10^{-5}$.

 Let us now discuss the experimental signatures for these processes. Due
to R-parity breaking, the slepton or the sneutrino which appears as
the decay product of the $W$ or $Z$ will itself decay predominantly into
quarks.
In the case of the slepton ( i.e.$\tilde {\tau}^+$ ), an analysis of the same
diagram as in Fig.1 shows that it decays predominantly to the
charm and strange quarks if sum the of the masses for the chargino and
neutralino is larger than that of the slepton as is usually expected in most
models. Turning
now to the sneutrino final state, it will predominantly decay into
$c\bar{c}$ and $b\bar{b}$ modes. Thus the overall signature for R-parity
violating radiative $W$ and $Z$ decays in this model will be two jets plus a
photon with the mass of the two-jets reconstructing to that of the
corresponding slepton. These two jets will having leading heavy flavor
components.
As far as feasibility of detection of the above processes is concerned,
the radiative $W$ decay suggested in this paper is probably beyond the reach
of LEPII but might be observable at a high luminosity $e^+e^-$ collider with
a larger center of mass energy. At hadron supercolliders such as the SSC or
LHC,
the rather large luminosity available implies[12] that we should expect more
than $10^3$ events per year. The main difficulty at such machines will be
the backgrounds from other sources. The radiative $Z$ decay should be
observable at LEPI by sitting on the $Z$ peak provided sufficient integrated
luminosity is accumulated.

 It is worth pointing out that in a non-supersymmetric two Higgs model,
if the charged Higgs boson is lighter than the $W$ boson, a similar appearing
decay can arise. It is however likely to be smaller due to the presence of
the vacuum mixing parameter $tan\beta$.

 \smallskip
{\bf\underline {IV.Neutralino Decay Modes}}

    In this section, we briefly mention some other distinct
experimental signatures of the model that arise in the neutralino
sector. This sector can be split into two parts in our scenario:
one heavy and the other light, the former consisting of
the $\nu_{\tau}^c$ and the heavy Zino and the latter consisting of
the Higgsinos and the lighter Zinos of the MSSM. The heavier neutralinos
can decay into $\tau^+ \tau^- \nu_{\tau}$, $\tau^- c\bar{s}$ etc.,
due to spontaneous R-parity breaking induced interactions.
In the light neutralino sector, if we ignore the small vev of
$\tilde{\nu}$, then there are only four mass eigenstates. While our
comments below apply to all the eigenstates which have a
significant photino component contained within them, we will give the generic
name photino to only one of the neutralinos in what follows.
A specific prediction of our model, in contrast to other
R-parity breaking models[13], is that the photino can decay only
to $\tau^- c \bar{s}$ as opposed to decays like $\tau^- \mu^+\nu_e$
etc. Such decay modes have recently been searched for by the OPAL
collaboration[14] with a null result. The present model on the other hand would
lead only to signatures of type $e^+e^-\to \tilde{\gamma}\tilde{\gamma}$
with the photinos subsequently decaying to $\tau^-c \bar{s}$.
This is a very different signature than previously considered.

{\bf\underline {V.Conclusion}}

   We have isolated a class of models with spontaneous R-parity
violation where the single photon radiative decays of $W$ and $Z$
are significant in contrast with the MSSM with explicit R-parity violation
wherein such decays are highly suppressed. Any evidence for such decays
could indicate the existence of new gauge symmetries beyond the
standard model. The attractive aspect of the model we consider here is
that in the limit of an
exact extended gauge symmetry, R-parity remains unbroken[7,15] so that any
manifestation of R-parity breaking at low energy appears only
in the form of lepton number violation and baryon number
remains an exactly conserved symmetry. Such models can appear
naturally in the low energy limit of some superstring theories.
This should provide a strong motivation to look for such
decays of the $W$ and $Z$ at LEP and at other colliders in the future.
We also point out some new R-parity breaking decays of the neutralinos
that can be sought at both LEPI and LEPII.

\begin{center}
{\bf Acknowledgement}
\end{center}

 One of us ( R.N.M.) would like to thank K.S.\ Babu and A.\ Jawahery
for discussions. T.G.R. would like to thank J.L.\ Hewett for discussions.
The work of R.N.M.was supported by a grant from
the National Science Foundation and the work of T.G.R. was
supported by a grant from the Department of Energy.

\newpage
%
\def\MPL #1 #2 #3 {Mod.~Phys.~Lett.~{\bf#1},\ #2 (#3)}
\def\NPB #1 #2 #3 {Nucl.~Phys.~{\bf#1},\ #2 (#3)}
\def\PLB #1 #2 #3 {Phys.~Lett.~{\bf#1},\ #2 (#3)}
\def\PR #1 #2 #3 {Phys.~Rep.~{\bf#1},\ #2 (#3)}
\def\PRD #1 #2 #3 {Phys.~Rev.~{\bf#1},\ #2 (#3)}
\def\PRL #1 #2 #3 {Phys.~Rev.~Lett.~{\bf#1},\ #2 (#3)}
\def\RMP #1 #2 #3 {Rev.~Mod.~Phys.~{\bf#1},\ #2 (#3)}
\def\ZP #1 #2 #3 {Z.~Phys.~{\bf#1},\ #2 (#3)}

\newpage

\noindent{\bf Figure Caption}
\begin{itemize}
\item[Figure 1.]{ The Feynman diagram that induces the R-violating
slepton-t-b coupling after B-L breakdown}
\item[Figure 2.]{ The one-loop diagram that induces the process
$W\rightarrow\tilde {\tau}\gamma$ decay amplitude.}
\item[Figure 3.]{ $W\rightarrow \tilde {\tau}\gamma$ branching fraction
predicted in our model as a function of the slepton mass assuming $f$=1. The
solid (dash-dotted, dashed) curve corresponds to a top-quark mass of 100
(150, 200)GeV.}
\item[Figure 4.]{$Z\rightarrow \snu\gamma$ branching fraction as a
function of the sneutrino mass assuming $f$=1. The curves are for the same
top-quark masses as in Fig.3.}
\end{itemize}


\begin{thebibliography}{99}
\bibitem{one} H.E.\ Haber and G.\ Kane, Phys. Rep. {\bf 117}, 75 (1985).
For a more recent overview of the situation, see X.Tata, Hawaii
Preprint (1991).
\bibitem{two} C.S.\ Aulakh and R.N.\ Mohapatra, Phys. Lett. {\bf B119},
136 (1983);
G.G.\ Ross and J.W.F.\ Valle, Phys. Lett. {\bf B151}, 375 (1985);
A.\ Santamaria and J.W.F.\ Valle, Phys. Lett. {\bf B195}, 423 (1987);
A.\ Masiero and J.W.F.\ Valle, Phys. Lett. {\bf B251}, 273 (1990).
\bibitem{three} L.\ Hall and M.\ Suzuki, Nucl. Phys. {\bf B231}, 419 (1984);
I.H.\ Lee, Nucl. Phys. {\bf B246}, 120 (1984);
S.\ Dawson, Nucl. Phys. {\bf B261}, 297 (1985);
R.\ Mohapatra, Phys. Rev. {\bf  D34}, 3457 (1986).
\bibitem{four} M.C.\ Gonzales-Garcia, J.C.\ Romao and J.W.F.\ Valle, FTUV-91-42
(1991); J.C.\ Romao, C.A.\ Santos and J.W.F.\ Valle, Phys. Lett. {\bf B288},
311 (1992);
V.\ Barger, G.F.\ Giudice and T.\ Han, Phys. Rev. {\bf D40}, 2987 (1989);
H.\ Dreiner and G.G.\ Ross, Nucl. Phys. {\bf B365}, 597 (1991);
D.P.\ Roy, CERN report (1992); R.M.\ Godbole, P.\ Roy and X.\ Tata, CERN
report CERN-TH.6613/92 (1992).
\bibitem{five} J.L.\ Hewett, Argonnne National Laboratory report
ANL-HEP-CP-92-23 (1992), and in the {\it Proceedings of the Workshop on Photon
Radiation From Quarks}, Annecy, France, December 2-3, 1991, CERN-Yellow-Report
92-04 (1992).
\bibitem{six} T.G.\ Rizzo, Argonne National Laboratory report
ANL-HEP-PR-92-81 (1992).
\bibitem{seven} R.N.\ Mohapatra,  Phys. Rev. {\bf D34}, 3457 (1986).
\bibitem{eight} R.N.\ Mohapatra, Phys. Rev. Lett. {\bf 56}, 561 (1986).
\bibitem{nine} See, for example, G.G.\ Ross and J.W.F.\ Valle in Ref.2.
\bibitem{ten} K.\ Yamamoto, Phys. Lett. {\bf B} (to appear).
\bibitem{eleven} J.F.\ Gunion, S.\ Dawson, H.\ Haber and G.\ Kane, {\it The
Higgs Hunter's Guide }, Addison-Wesley (1990).
\bibitem{twelve} E.\ Eichten, I.\ Hinchliffe, K.\ Lane, and C.\ Quigg,
\RMP 56 579 1984 .
\bibitem{thirteen} K.S.\ Babu and R.N.\ Mohapatra, Phys. Rev. {\bf D42}, 3778
(1991).
\bibitem{fourteen} C.Y.\ Chang and G.\ Snow, for the OPAL collaboration (to
appear as CERN preprint).
\bibitem{fifteen} T.G.\ Rizzo, Argonne National Laboratory report
ANL-HEP-PR-92-95 (1992), to appear in Phys. Rev. {\bf D}.

\end{thebibliography}
\end{document}